\documentclass[prl,reprint,twocolumn,showpacs,superscriptaddress,floatfix,aps]{revtex4-2}

\usepackage{amssymb,amsmath,latexsym,bm,dsfont,graphicx,braket}
\usepackage[T1]{fontenc} 
\usepackage{txfonts}
\graphicspath{{../../figures/}{./}} 
\makeatletter\def\@pacs@name{DOI: }\makeatother 

\renewcommand{\vec}[1]{{\bm{\mathrm{#1}}}}
\newcommand{\vhat}[1]{\hat{\bm{\mathrm{#1}}}}
\let\epsilon\varepsilon
\let\Re\undefined
\let\Im\undefined
\DeclareMathOperator{\Im}{Im}
\DeclareMathOperator{\Re}{Re}

\renewcommand{\tilde}{\widetilde}

\usepackage{xcolor}
\usepackage[normalem]{ulem} 

\usepackage[colorlinks=true, citecolor=blue, linkcolor=blue, urlcolor=blue]{hyperref} 

\begin{document}
\title{Electronic Orbital Angular Momentum Driven by Finite-Momentum Phonons:\\Beyond Chiral and Axial Phonons}
\author{Yongho Park}%
\affiliation{Department of Physics, Yonsei University, Seoul 03722, Korea}%
\affiliation{Center for Quantum Dynamics of Angular Momentum, Pohang University of Science and Technology, Pohang 37673, Korea}%
\author{Jeonghun Sohn}%
\email{probably@postech.ac.kr}
\affiliation{Department of Physics, Pohang University of Science and Technology, Pohang 37673, Korea}%
\affiliation{Center for Quantum Dynamics of Angular Momentum, Pohang University of Science and Technology, Pohang 37673, Korea}%
\author{Sejoong Kim}%
\email{sejoong@hongik.ac.kr}
\affiliation{Department of Electronic and Electrical Convergence Engineering, Hongik University, Sejong 30016, Korea}%
\affiliation{Center for Quantum Dynamics of Angular Momentum, Pohang University of Science and Technology, Pohang 37673, Korea}%
\author{Kyoung-Whan Kim}%
\email{kwkim@yonsei.ac.kr}
\affiliation{Department of Physics, Yonsei University, Seoul 03722, Korea}%
\affiliation{Center for Quantum Dynamics of Angular Momentum, Pohang University of Science and Technology, Pohang 37673, Korea}%
\date{\today}

\begin{abstract}
We show that finite-momentum phonons generate electronic orbital angular momentum (OAM) even without phonon axiality or chirality, with the response controlled by the phonon wave vector $q$ and frequency $\omega$. We develop a general gauge-field theory in which a unitary transformation absorbs the phonon displacement into emergent vector and scalar potentials acting on the electrons, providing a perturbative classification of the OAM response in $q$ and $\omega$. 
We derive $q$- and $\omega$-scaling laws for AC and DC responses. Notably, AC OAM arises even for linearly polarized phonons, with its magnitude and sign tunable by $q$. Its generation is governed by the matching between the phonon geometry and the electronic orbital texture rather than solely by the phonon angular momentum. The connection of the DC response to the electronic Berry curvature further supports that phonon angular momentum is not the only degree of freedom governing electronic OAM generation. Time-dependent tight-binding simulations 
under acoustic-phonon driving independently confirm the core predictions. Our results extend phonon-driven OAM beyond chiral and axial phonons and establish a wave-vector-tunable route to orbitronics, accessible with surface acoustic waves.
\end{abstract}


\maketitle
\section{I.\quad Introduction} A circularly polarized phonon carries angular momentum. This phonon angular momentum (PAM) has become an actively studied resource, linked to the Einstein--de Haas
effect~\cite{zhangAngularMomentumPhonons2014,ruckriegelAngularMomentumConservation2020}, the phonon magnetic moment~\cite{juraschekOrbitalMagneticMoments2019,renPhononMagneticMoment2021},
dynamical multiferroicity and the phonon inverse Faraday effect~\cite{geilhufeDynamicallyInducedMagnetism2021,basiniTerahertzElectricfielddrivenDynamical2024,shabalaPhononInverseFaraday2024}, and spin/orbit--lattice coupling~\cite{ruckriegelAngularMomentumConservation2020,chaudharyGiantEffectiveMagnetic2024b}. Magnetic and orbital responses mediated by electron--phonon coupling are also being explored in Dirac, topological, and magnetic systems~\cite{chengLargeEffectivePhonon2020,baydinMagneticControlSoft2022,gurtubayMagneticOscillationsInduced2020,chenGaugeTheoryGiant2025,chenGeometricOriginPhonon2025,xueExtrinsicMechanismsPhonon2025}. More broadly, the development of orbitronics is opening a new direction that, beyond PAM itself, exploits the electron--phonon angular-momentum coupling~\cite{yaoDynamicalOrbitalAngular2025a,satoOrbitalAccumulationInduced2025} as a route to control the electronic state 
and to detect phonons.

Along these lines, most studies have focused on chiral or axial phonons~\cite{zhangChiralPhononsHighSymmetry2015,zhuObservationChiralPhonons2018,ishitoTrulyChiralPhonons2023,suriChiralPhononsMoire2021,komiyamaPhysicsPhononsSystems2022,bendinDWavePhononAngular2025,sunPhononAngularMomentum2025}
, under the common assumption that a finite PAM is required to generate an electronic orbital angular momentum (OAM) response. Moreover, recent studies have largely been restricted to specific settings involving chiral (axial) phonons~\cite{yaoDynamicalOrbitalAngular2025a,satoOrbitalAccumulationInduced2025,pezoFirstprinciplesPredictionChiralphononinduced2026a} or surface acoustic waves~\cite{taniguchiAcousticGenerationOrbital2025a,wuAcousticOrbitalHall}, leaving a general treatment of electron--phonon angular-momentum coupling still lacking. A general framework would enable a systematic understanding of the role of PAM across different response channels, including the AC and DC components and the local and global contributions, while establishing their distinct dependence on the phonon wave vector $q$ and frequency $\omega$. Such a classification may also provide experimentally accessible signatures for distinguishing the different contributions.



In this paper, we analyze the phonon-induced electronic OAM response within a gauge-field framework. In this framework, the effects of phonons are interpreted as effective vector and scalar potentials acting on electrons, which are proportional to the phonon wave vector $q$ and the phonon frequency $\omega$, respectively. This formulation allows us to systematically separate the AC and DC response channels and establish their $q$- and $\omega$-scaling within perturbation theory. The resulting scaling relations are independently confirmed by our time-dependent simulations, based on tight-binding models~\cite{hamada2020PRR,yao2022PRB,yao2025PRB,hanOrbitalPumpingIncorporating2025a,yaoDynamicalOrbitalAngular2025a}.

Beyond providing a general and intuitive description of the phonon-induced electronic responses, this framework clearly reveals how fundamental solid-state quantities such as the Berry curvature, may enter the resulting phonon-driven electronic response, while identifying $q$ and $\omega$ as distinct parameters controlling the response. Most importantly, we find that electronic OAM can be generated even by a linearly polarized phonon (neither chiral nor axial), with both its magnitude and sign controllable by $q$. Thus, phonon-induced electronic OAM is not restricted to chiral or axial phonons, extending OAM generation beyond a picture based solely on PAM.

\section{II.\quad Phonon-induced gauge fields} To model an electronic Hamiltonian in the presence of phonons, we start from the Hamiltonian of an electron in a deformable lattice,
\begin{equation}
  H=\frac{\mathbf p^2}{2m_e}+V\left(\mathbf r-\mathbf u(\mathbf r,t)\right),
\end{equation}
where $\mathbf p$ is the electron momentum, $m_e$ is the electron mass,
$V$ is the crystal potential, and $\mathbf u(\mathbf r,t)$ is the phonon
displacement field. Throughout, we treat the displacement $\mathbf u$
as small and expand in powers of it. The extension to higher orders is
straightforward.
We also take $\mathbf u(\mathbf r,t)$ as a slowly varying continuous field,
which describes acoustic phonons. 
In treatments that neglect the finite phonon wave vector~\cite{yaoDynamicalOrbitalAngular2025a}, nonuniform ionic displacements are typically introduced through optical modes, which generally lie at higher energies. By retaining the spatial variation of $\vec{u}(\vec{r},t)$, our framework instead allows such effects to arise from low-energy acoustic phonons. 
Optical phonons are captured by assigning
a separate displacement to each sublattice, a generalization we leave to
future work.

To absorb the lattice displacement into the electronic frame, we introduce a unitary
transformation $U=e^{-iG/\hbar}$, where $G$ is expanded in powers of $\vec{u}$, $G=G^{(1)}+G^{(2)}+\cdots$ and the leading-order generator is $G^{(1)}=\frac12(\mathbf p\cdot\mathbf u+\mathbf u\cdot\mathbf p)$, whose justification is given in Supplemental Material~\cite{supple}. 
After the transformation, the space-time dependence of $\vec{u}(\vec{r},t)$ leads to the following minimal-coupling form for the transformed Hamiltonian:
\begin{equation}
  H'=\frac{(\mathbf p-e\mathbf A)^2}{2m_e}+V(\mathbf r)+e\phi,
\end{equation}
where the emergent gauge potentials are given by $e\mathbf A=i\hbar\,U^\dagger\nabla U$ and $e\phi=-i\hbar\,U^\dagger\partial_t U$. The emergent vector potential $e\mathbf A$ arises from the spatial variation of the displacement field, whereas the emergent scalar potential $e\phi$ originates from its time dependence.~\cite{footnote1} Analogous strain-induced gauge fields and synthetic electric fields have been discussed in graphene and carbon nanotubes~\cite{vonoppenSyntheticElectricFields2009,sohierPhononlimitedResistivityGraphene2014}.
In this way, an electron in the presence of phonons can be described as an electron on a static lattice subject to phonon-induced vector and scalar potentials, which can be treated as \textit{perturbations to the phonon-free electronic system}. Its orbital response can be classified according to the perturbation channels $H^{(1)},H^{(2)},\ldots$ introduced below. An important advantage of this formulation is that, to leading order in the slowly varying approximation, the space-time dependence of the phonon fields enters parametrically, as in standard semiclassical wave-packet treatments of slowly varying perturbations~\cite{xiaoBerryPhaseEffects2010a}
, greatly simplifying the analysis while retaining a transparent physical interpretation.

Expanding $e\mathbf A$ and $e\phi$ order by order in the displacement field, the transformed Hamiltonian becomes $H'=H_0+H^{(1)}+H^{(2)}+\cdots$, where $H_0$ describes Bloch electrons on the undistorted lattice. The order counting is transparent for a single phonon mode,
$
  \mathbf u(\mathbf r,t)=\mathbf u_0\,e^{i(\mathbf q\cdot\mathbf r-\omega t)}+\mathrm{c.c.},
$
where $\mathbf q$ is the phonon wave vector with magnitude $q$, $\omega$ the phonon frequency, and $u_0\equiv|\mathbf u_0|$ the displacement amplitude. The first-order potentials then follow as
$e\mathbf A^{(1)}=\nabla G^{(1)}$ and $e\phi^{(1)}=-\partial_t G^{(1)}$. The first-order vector potential $e\mathbf A^{(1)}\propto q$ originates from the spatial gradient $\partial_i u_j$, whereas the scalar potential $e\phi^{(1)}\propto\omega$ originates from the time derivative $\dot{\mathbf u}$. Phonon-induced electronic responses can be calculated by using perturbation theory. Contributions that are first order in $\vec{u}(\vec{r},t)$ give rise to oscillating responses (AC), whereas second-order contributions can contain a DC component because the time average of $u_i(\vec{r},t)u_j(\vec{r},t)$ need not vanish~\cite{footnote2}.


We target the orbital response that a time-dependent, spatially varying phonon induces in the electronic orbital structure rather than a uniform rigid translation or a static distortion, so the $q=0$ contribution is excluded. By contrast, the $q$-dependent terms are kept, since they describe a relative deformation or a
phonon-induced gauge field within the lattice. Likewise, the $\omega=0$ terms correspond to a static distortion. This can change the electronic structure, but it differs from the time-dependent phonon-induced response of interest. Thus we focus on the $\omega$-dependent dynamical contribution that arises from the time-varying displacement field. 


\section{III.\quad AC orbital response}
We calculate the AC orbital response 
to first order in the phonon-induced perturbation. In the gauge-field formulation,
the first-order perturbation reads
\begin{equation}
  H^{(1)}=-\frac{1}{2m_e}\{\mathbf p,e\mathbf A^{(1)}\}+e\phi^{(1)} .
  \label{eq:H1}
\end{equation}
Here $e\phi^{(1)}\propto\omega$ carries the explicit
time dependence, whereas $-\{\mathbf p,e\mathbf A^{(1)}\}/2m_e$ acts as a spatial-gradient vertex, with its time dependence entering through the phonon phase. We therefore concentrate on the $e\phi^{(1)}$ contribution to isolate the $\omega$-dependent dynamical AC response that scales explicitly with $\omega$, treating the vector-potential term as a spatial-gradient sector. In the lowest-order adiabatic approximation, time enters parametrically, allowing us to apply time-independent perturbation theory at each instant. We then obtain the perturbed eigenstate $|\tilde u_{n\mathbf k}\rangle=|n\mathbf k\rangle +\sum_{m\mathbf k'\neq n\mathbf k}|m\mathbf k'\rangle\, \langle m\mathbf k'|H^{(1)}|n\mathbf k\rangle/ (E_{n\mathbf k}-E_{m\mathbf k'})$. The expectation value of an observable $A$ then follows from the statistical average $\langle A\rangle=\sum_{n\mathbf k}f_{n\mathbf k}\langle\tilde u_{n\mathbf k}|A|\tilde u_{n\mathbf k}\rangle$, where $f_{n\vec{k}}$ is the electronic distribution function. The observable $A$ itself is not transformed by $U$, so that the resulting OAM is referenced to the instantaneous atomic centers. Transforming the operator would reintroduce the rigid lattice motion, a mechanical contribution rather than a local electronic orbital response, consistent with our tight-binding calculation. Writing $A_{n\mathbf k\,m\mathbf k'}\equiv\langle n\mathbf k|A|m\mathbf k'\rangle$ and analogously for other operators, the $\phi^{(1)}$ contribution to the AC component of an observable $A$ (such as OAM and OAM current) reads
\begin{equation}
  \langle A\rangle_{\rm AC}(t)=2\Re
  \sum_{m\mathbf k'\neq n\mathbf k}
 f_{n\mathbf k} \frac{A_{n\mathbf k\,m\mathbf k'}\,
        e\phi^{(1)}_{m\mathbf k'\,n\mathbf k}}
       {E_{n\mathbf k}-E_{m\mathbf k'}} . \label{eq:AC}
\end{equation}
The scalar potential $e\phi^{(1)}$ contains the phase factor $e^{i\mathbf q\cdot\mathbf r}$, which connects a state at $\mathbf k$ to states at $\mathbf k'=\mathbf k\pm\mathbf q$ and thereby transfers momentum between the electron and the phonon. Evaluating the response then requires the $\mathbf k$-off-diagonal matrix elements of the observable $A$.

Within the slowly varying approximation, however, it is natural to recast the problem in a local Bloch description~\cite{kimIntrinsicSpinTorque2015} by decomposing the electronic coordinate as $\mathbf r=\mathbf R+\mathbf r_{\rm cell}$. Here $\mathbf R$ is the macroscopic coordinate over which the phonon field varies slowly, and $\mathbf r_{\rm cell}$ describes the orbital structure within the unit cell. In the long-wavelength limit, we retain the macroscopic phase and expand only the intracell factor as $e^{i\mathbf q\cdot\mathbf r}\approx e^{i\mathbf q\cdot\mathbf R}(1+i\mathbf q\cdot\mathbf r_{\rm cell})$. At fixed $\vec{R}$, the macroscopic phase $e^{i\vec{q}\cdot\vec{R}}$ is treated parametrically in the local electronic problem~\cite{xiaoBerryPhaseEffects2010a}, whereas $e^{i\mathbf q\cdot\mathbf r_{\rm cell}}$ retains the intracell spatial dependence. The response can then be expressed in terms of $k$-diagonal matrix elements of the observable. As discussed above, we discard the zeroth-order contribution in $q$, which does not generate a relative orbital response of interest here. Therefore, the leading nontrivial AC contribution comes from the $i\mathbf q\cdot\mathbf r_{\rm cell}$ term. With this treatment, Eq.~(\ref{eq:AC}) reduces to the following $\vec{k}$-diagonal form:
\begin{equation}
  \langle A\rangle_{\rm AC}(t)=
  -\omega\Re\sum_{m\ne n,\vec{k}}
    \frac{f_{nm,\mathbf k}}{E_{nm,\mathbf k}}\,A_{nm,\mathbf k}\,
    [(\mathbf q\cdot\mathbf r_{\rm cell})
    (\mathbf u(\mathbf R,t)\cdot\mathbf p)]_{mn,\mathbf k},
  \label{eq:ACleading}
\end{equation}
where $f_{nm,\mathbf k}\equiv f_{n\mathbf k}-f_{m\mathbf k}$,  $E_{nm,\mathbf k}\equiv E_{n\mathbf k}-E_{m\mathbf k}$, and $[\cdots]_{mn,\mathbf k}\equiv\langle m\mathbf k|\cdots|n\mathbf k\rangle$ for operators. Equation~\eqref{eq:ACleading} shows that the leading finite-$q$ AC response has an explicit $q\omega$ scaling and depends on the electronic interband structure as well as on the phonon geometry. The $\mathbf q\cdot\mathbf r_{\rm cell}$ reflects the weak variation of the phonon phase within the unit cell and thereby probes the electronic intracell orbital structure. In turn, $\mathbf u(t)\cdot\mathbf p$ couples the lattice-displacement direction to the electronic momentum matrix element.

\begin{figure}[b]
	\includegraphics[width=\linewidth]{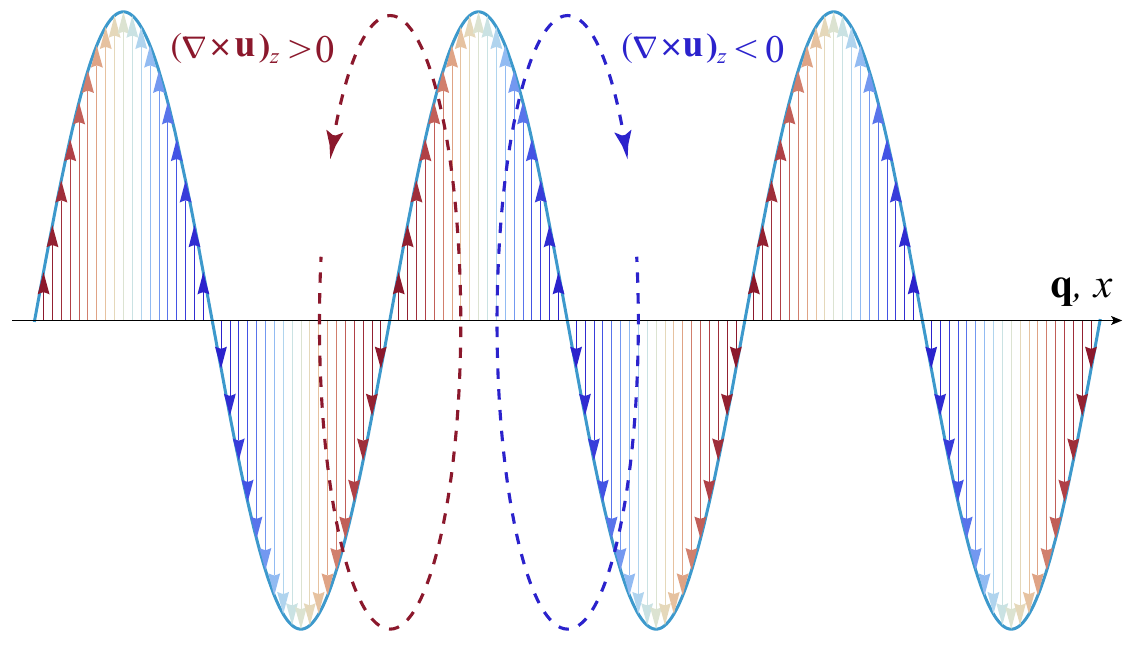}
	\caption{Local generation of orbital angular momentum by the vorticity of a linearly polarized phonon.}
	\label{fig:OAM by linear pol}
\end{figure}

Replacing $A$ by the OAM operator $\vec{L}$ gives the AC expression for the phonon-induced OAM. According to Eq.~(\ref{eq:ACleading}), the AC orbital polarization is governed by the phonon propagation direction, polarization direction, and orbital texture of the Bloch state, rather than by PAM alone. 
%
An OAM component can arise whenever allowed by the symmetry of Eq.~(\ref{eq:ACleading}), without requiring finite PAM. An interesting example follows by considering the Hermitian part of the vertex in Eq.~(\ref{eq:ACleading}), $\frac{1}{2}\{\vec{q}\cdot\vec{r}_{\rm cell},\vec{u}\cdot\vec{p}\}$, which can be decomposed into symmetric and antisymmetric parts as $\frac{1}{4}(q_iu_j+q_ju_i)\{r_{\rm cell, \it i},p_j\}+\frac{1}{2}(\vec{q}\times\vec{u})\cdot\vec{L}$, where the Einstein summation convention is used and $\vec{L}=\vec{r}_{\rm cell}\times\vec{p}$. The second contribution provides a direct symmetry-allowed channel for generating $L_z$ when $\vec{q}\parallel\vhat{x}$ and $\vec{u}\parallel\vhat{y}$, demonstrating that \textit{a linearly polarized phonon can generate OAM}. This shows that OAM generation does not require the phonon mode itself to be axial or chiral~\cite{footnote3}. 
The relevant axial structure instead originates from the combined propagation and polarization structure of the phonon, characterized by the nonzero phonon vorticity $\nabla\times\vec{u}$ depicted in Fig.~\ref{fig:OAM by linear pol}, rather than from PAM, $\mathbf J_{\rm ph}\propto \vec{u}\times\dot{\vec{u}}$~\cite{zhangAngularMomentumPhonons2014}.
This is explicitly demonstrated by the numerical simulations presented below. A possible experimental scheme for detecting this OAM is also discussed later in this paper. As a remark, the symmetric part of the vertex is related to strain and can also generate OAM, implying that a finite vorticity is not a necessary condition for an AC orbital response.

\section{IV.\quad DC orbital response}
The DC orbital response has two sources, which we call DC1 and DC2. DC1 originates from the second-order scalar potential $e\phi^{(2)}$, and DC2 from the cross term between the first-order potentials $e\mathbf A^{(1)}$ and $e\phi^{(1)}$. The other terms are static ($\omega=0$), rigid ($q=0$), or higher order in the slowly varying regime, such as the $e^2\vec{A}^{(1)}\cdot\vec{A}^{(1)}/2m_e$ term. We first consider DC1. Since $e\phi^{(2)}$ is second order in the phonon
amplitude $u_0$, it can be treated within first-order perturbation theory.

The linearity of first-order perturbation theory allows us to replace $e\phi^{(2)}$ by its time average $e\phi^{(2)}_{\rm DC}\equiv \braket{e\phi^{(2)}}_t$, before applying the perturbation formula. 
From the expression for $e\phi^{(2)}$~\cite{supple}, we obtain $e\phi^{(2)}_{\rm DC}=(2i\hbar)^{-1}\braket{[G^{(1)},\dot{G}^{(1)}]}_t$, which gives rise to a contribution proportional to $q\omega$.
Here, the DC component arises from the $e^{i(\mathbf q\cdot\mathbf r-\omega t)}\times e^{-i(\mathbf q\cdot\mathbf r-\omega t)}$ combination
after time averaging. In this combination, the spatial phase factors also cancel out and thus
$e\phi^{(2)}_{\rm DC}$ does not change the electronic crystal momentum. Thus
the DC1 response can be expressed in terms of interband matrix elements between different bands at the same $\vec{k}$.

Applying first-order perturbation theory to $e\phi^{(2)}_{\rm DC}$ gives a phonon-induced DC correction to a physical observable $A$.
\begin{equation}
  \langle A\rangle_{\rm DC1}=-2\omega \Re[(\mathbf u_0^*\cdot
  \mathbf q)\mathbf u_0]\cdot\Re\sum_{m\ne n,\vec{k}}
  \frac{f_{nm,\mathbf k}}{E_{nm,\mathbf k}}\,A_{nm,\mathbf k}\,
  \mathbf p_{mn,\mathbf k}.\label{eq:DC1}
\end{equation}
The factor $-2\omega\Re[(\mathbf u_0^*\cdot \mathbf q)\mathbf u_0]$ is determined by the phonon part, whereas the summation reflects the electronic part, emphasizing their combined roles rather than PAM alone. As an illustration, for a linearly polarized phonon, $-2\omega\Re[(\mathbf u_0^*\cdot
\mathbf q)\mathbf u_0]=-2u_0^2(\vhat{e}\cdot\vec{q})\cdot \vhat{e}\omega $, where $u_0$ is the phonon amplitude and $\vhat{e}$ is the polarization direction. For a circularly polarized phonon $-2\omega\Re[(\mathbf u_0^*\cdot
\mathbf q)\mathbf u_0]=-u_0^2\vec{q}_\parallel\omega$, where $\vec{q}_\parallel$ is the phonon momentum vector projected onto the circular polarization plane. The overall scaling relation of the DC1 contribution is thus $q\omega$.

A further important feature of Eq.~\eqref{eq:DC1} lies in its electronic part. For instance, setting $A=L_\alpha$, the integrand of the DC1 response is 
odd under $\mathbf k\to-\mathbf k$ due to inversion symmetry, so that the Brillouin-zone sum vanishes
. By contrast, in an inversion-breaking crystal, a chiral crystal, or an interface geometry, a lower-order orbital response may be allowed~\cite{hamadaPhononAngularMomentum2018,hamadaPhononRotoelectricEffect2020}. In this respect the DC1 channel does not require a finite phonon angular momentum and can arise even for a linearly polarized phonon once inversion symmetry is broken. 
On the other hand, a DC orbital current $\{L_\alpha,\vec{v}\}$ is instead inversion-odd, so its integrand is even under $\mathbf k\to-\mathbf k$ and it can be generated even in an inversion-symmetric system. The electronic part also reveals an interesting connection for a different choice of $A$. When $A_{nm,\mathbf k}$ is replaced by the intracell position operator $r_{i,nm,\mathbf k}=-i\hbar v_{i,nm,\mathbf k}/E_{nm,\mathbf k}$, where $\vec{v}$ is the velocity operator, the electronic part reads $-(m_e/\hbar)\vec{\Omega}_i$ where
\begin{equation}
	\vec{\Omega}_i=-\hbar^2 \Im\sum_{m\ne n,\vec{k}}
	f_{nm,\mathbf k}\frac{v_{i,nm,\mathbf k} \vec{v}_{mn,\mathbf k}}{E_{nm,\mathbf k}^2},
\end{equation}
which is precisely the Berry curvature~\cite{xiaoBerryPhaseEffects2010a}, summed over the occupied states. For instance, a linearly polarized phonon along $y$ ($q_y\ne 0$) induces an $x$-polarized electronic response in a system with finite value of $\Omega_{xy}$. 
This example suggests that the gauge-field framework extends beyond OAM to a broader scope, including phonon-induced responses governed by geometric electronic properties such as the Berry curvature, which may thereby be probed through electronic responses.

We now turn to the second channel, DC2. Apart from the scalar-potential channel, a DC response also arises from the second-order response to the first-order phonon-induced gauge perturbations, the $H^{(1)}H^{(1)}$ channel. It is the DC component remaining after the product of two oscillating first-order vertices is time-averaged, and is therefore second order in the phonon amplitude.
\begin{equation}
  \langle A\rangle_{\rm DC}^{(2)}=\!\!\sum_{n,m,l,\mathbf k,\mathbf k'}\!\!
  \mathcal F_{nml}(\mathbf k,\mathbf k')\,A_{nm,\mathbf k}\,
  H^{(1)}_{m\mathbf k\,l\mathbf k'}\,H^{(1)}_{l\mathbf k'\,n\mathbf k},
\end{equation}
where $\mathcal F_{nml}$ collects the Fermi factors and energy denominators, with its full form given in the Supplemental Material~\cite{supple}. A genuine DC response requires both opposite-phase factors $e^{\pm i\mathbf q\cdot\mathbf r}$ to enter, one from each vertex. These factors enforce momentum conservation, so the leading vertex product describes a virtual process in which the electron is scattered $\mathbf k\to\mathbf k\pm \mathbf q\to\mathbf k$ by the phonon, with the net crystal momentum conserved but the intermediate state shifted by $\mathbf q$. At zeroth order in the $q$-expansion of the intermediate state, the intermediate state returns to $k$. In an inversion-symmetric system where the DC1 contribution vanishes by inversion parity, the corresponding $q\omega$ contribution also vanishes by the same argument. The leading nonzero DC2 response then requires expanding the intermediate state, matrix element, or denominator to first order in $q$, which, combined with the factor of $q$ already present, yields a $q^2\omega$ contribution. Due to the complexity, the first-order vertices and the full vertex product are given in the Supplemental Material~\cite{supple}, and we revisit the scaling law in the independent numerical calculations shown below. The $q^2\omega$ response is odd under the phonon helicity, linking the DC2 orbital moment to a finite phonon angular momentum. A helicity-independent moment would instead require additional symmetry breaking or spatial inhomogeneity. More detailed studies of DC2 contributions are left for future work.


\section{V.\quad Numerical comparison with tight-binding formalism}
Previously, phonon-induced electronic responses have been studied using tight-binding models~\cite{hamada2020PRR,yao2022PRB,yao2025PRB,hanOrbitalPumpingIncorporating2025a,yaoDynamicalOrbitalAngular2025a}. In the presence of lattice displacements, the bond lengths and orientations between neighboring atoms vary dynamically, leading to perturbative corrections to the electronic Hamiltonian. Here, we use such a microscopic tight-binding construction, independent of the gauge-field derivation, to test the predicted scaling laws and explicitly demonstrate electronic orbital angular momentum induced by linearly polarized phonons.

We consider a quasi-two-dimensional tight-binding model with $p$ orbitals on a square lattice, whose explicit Hamiltonian and model details are presented in the Appendix. The phonon effects are implemented through spatiotemporally varying hopping parameters, and the time-dependent Schrödinger equation is solved using the tkwant package~\cite{klossTkwantSoftwarePackage2021}. 
We focus on low-energy acoustic phonons, for which surface acoustic waves provide a possible experimental realization and have recently emerged as a platform for orbital-current generation~\cite{taniguchiAcousticGenerationOrbital2025a,rovirolaChiralphononGenerationOrbital2025,wuAcousticOrbitalHall}. In this low-energy regime, the slowly varying lattice displacement also connects naturally to an adiabatic pumping picture~\cite{thoulessQuantizationParticleTransport1983}, while extensions to optical and more general phonon modes are left for future work. 

\begin{figure}
  \centering
  \includegraphics[width=\linewidth]{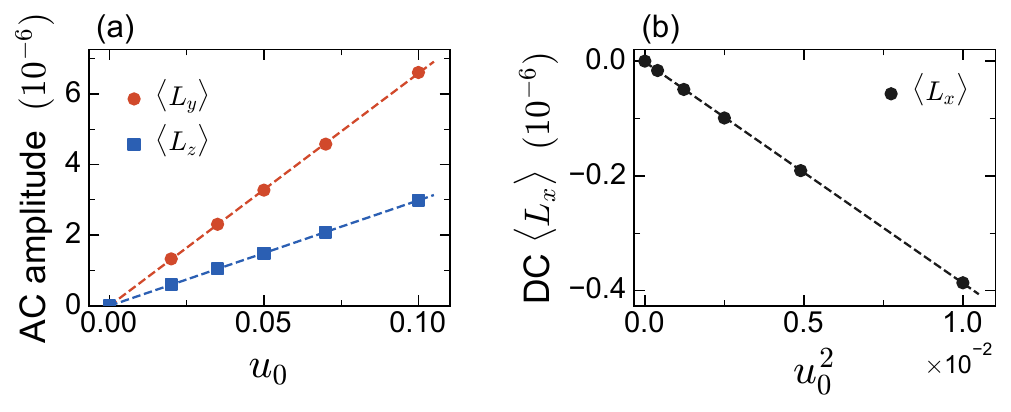}
  \caption{Phonon-amplitude scaling for circularly polarized phonons in the $yz$ plane. (a) The AC orbital amplitude is
  linear in $u_0$ ($\propto u_0$). (b) The time-averaged DC component of
  $\langle L_x\rangle$ is quadratic in $u_0$ ($\propto u_0^2$). Here $u_0$ is the phonon amplitude.}
  \label{fig:Ju0}
\end{figure}

We first check that the AC and DC channels exhibit distinct amplitude scalings with $u_0$. We apply a circularly polarized phonon in the $yz$ plane, $\mathbf u=(0,\,u_0\cos(qx-\omega t),\,u_0\sin(qx-\omega t))$, and calculate the AC amplitude and DC contribution, with the latter obtained by time averaging. 
Figure~\ref{fig:Ju0} shows that the AC amplitude is linear in $u_0$, consistent with first-order perturbation theory, whereas the DC component is quadratic in $u_0$, consistent with the rectification of two first-order vertices. These scalings confirm that the AC and DC signals originate at the perturbative orders identified analytically.

\begin{figure}[b]
  \centering
  \includegraphics[width=\linewidth]{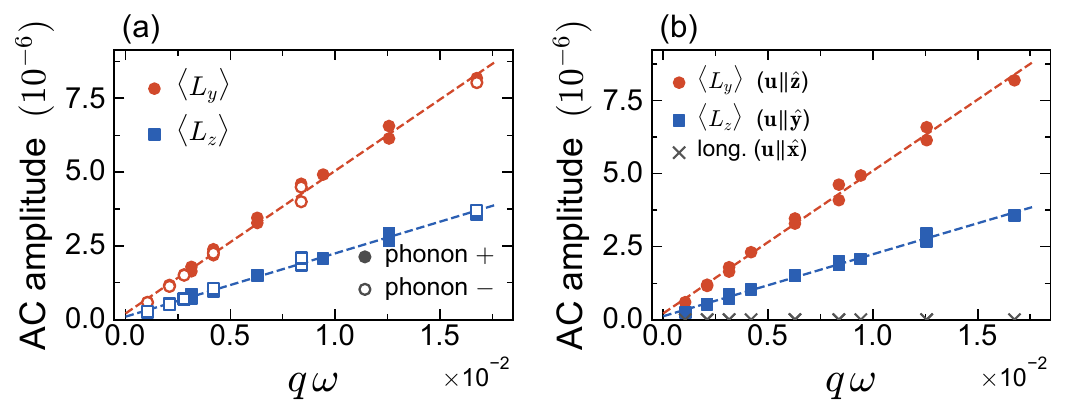}
  \caption{$q\omega$ scaling of the AC orbital amplitude. (a) A $yz$-circular phonon; filled/open markers denote the phonon helicity. (b) Linearly polarized phonons: transverse
  ($\mathbf u\perp\mathbf q$) induce $\langle L_y\rangle,\langle L_z\rangle$,
  while longitudinal ($\mathbf u\parallel\mathbf q$, $\times$) induce none.}
  \label{fig:ACqw}
\end{figure}

To examine the leading dynamical scaling of the AC orbital
response, we plot the AC amplitude of the $yz$-circular phonon, which appears in the
$L_y$ and $L_z$ components rather than in $L_x$, against $q\omega$.
Figure~\ref{fig:ACqw}(a) shows a linear relation with
high fitting quality~\cite{footnote4}. 
The suppression of the $L_x,L_y$ AC components in the
$xy$-phonon case reflects the weak $p_z$ weight at the chosen Fermi level. Related phonon-selective orbital responses have also been reported in chiral-phonon current, orbital-current, and orbital-Seebeck phenomena~\cite{rovirolaChiralphononGenerationOrbital2025,nabeiOrbitalSeebeckEffect2026,taniguchiAcousticGenerationOrbital2025a}.

To examine OAM induced by linearly polarized phonons, we consider linearly polarized components along $\vhat{y}$ and $\vhat{z}$. Figure~\ref{fig:ACqw}(b) shows that a transverse linear phonon ($\mathbf u\perp\mathbf q$) alone produces a finite AC OAM. A phonon with $\mathbf u\parallel\vhat y$ induces $\langle L_z\rangle$ and a phonon with $\mathbf u\parallel\vhat z$ induces $\langle L_y\rangle$, both linear in
$q\omega$. The similar magnitudes of red and blue lines in Figs.~\ref{fig:ACqw}(a) and \ref{fig:ACqw}(b) imply that $L_y$ and $L_z$ AC responses of the $yz$-circular phonon are given by the superposition of the two transverse linearly polarized contributions. A longitudinal linear phonon ($\mathbf u\parallel\mathbf q\parallel\vhat x$), by contrast, produces no OAM, since it only stretches the bond ($\Delta\mathbf d^{\perp}=0$) and induces no orbital mixing, which can also be understood from $\nabla\times \vec{u}=0$. The symmetric strain part of the vertex remains finite in this geometry, but its contribution is canceled by the mirror symmetry of the square lattice, again consistent with the absence of orbital mixing. 
The AC OAM induced in the transverse geometry is a local oscillating orbital polarization rather than a net angular-momentum generation, as the response averages to zero over one phonon wavelength. Even so, a phonon propagating into an adjacent magnet can still exert an AC orbital torque, which offers an experimental probe. 

\begin{figure}
  \centering
  \includegraphics[width=\linewidth]{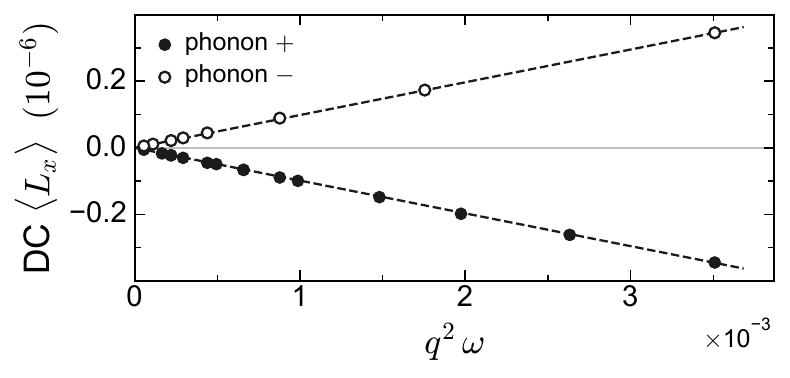}
  \caption{$q^2\omega$ scaling of the DC component of
  $\langle L_x\rangle$ in the presence of a circularly polarized phonon in the $yz$ plane. Filled and open markers denote $+$ and $-$ phonon
  helicity.
  }
  \label{fig:q2w}
\end{figure}

Lastly, we show the scaling of the DC component, which is expected to follow $q^2\omega$ in a centrosymmetric system.
Figure~\ref{fig:q2w} shows a linear dependence on $\propto q^2\omega$, in agreement with our expectation. The  high fitting quality and the helicity-dependent sign reversal support that the DC signal is not a numerical offset or fitting artifact but a genuine rectified response. 
Experimentally, this DC orbital accumulation corresponds to a static OAM generated under phonon drive, detectable through a DC orbital torque on an adjacent ferromagnet~\cite{leeOrbitalTorqueMagnetic2021} or through the magneto-optical Kerr effect~\cite{choiObservationOrbitalHall2023,lyalinMagnetoOpticalDetectionOrbital2023}.

\section{IV.\quad Conclusion}
In summary, by absorbing the phonon displacement through a unitary
transformation, we developed a gauge-field formulation that classifies
phonon-induced electronic OAM into AC and DC channels and identifies
their $(q,\omega)$ scaling. The leading dynamical AC response scales as
$q\omega$, while the rectified DC response scales as $q^2\omega$ once
lower-order contributions associated with symmetry, the static limit,
and rigid translation are excluded. The response is governed by the
finite-$q$ phonon geometry and the electronic orbital texture rather
than by PAM alone: even a linearly polarized phonon with
$\mathbf J_{\rm ph}=0$ can generate OAM, with its magnitude and sign
controlled by $q$. This extends phonon-induced OAM beyond chiral and
axial phonons.

The resulting AC orbital accumulation may enable dynamical control of
magnetization through an orbital torque at a ferromagnetic
junction~\cite{leeOrbitalTorqueMagnetic2021}, while surface acoustic
waves~\cite{taniguchiAcousticGenerationOrbital2025a,rovirolaChiralphononGenerationOrbital2025,wuAcousticOrbitalHall}
offer a natural platform in which $q$ and $\omega$ can be tuned
directly. More broadly, acoustic-phonon-induced orbital responses may
provide an additional ingredient in understanding orbital relaxation
and transport in
metals~\cite{fukami2026,liaoLongRangeVersusLocal2026,giustinoElectronphononInteractions2017}.
Extensions to optical phonons, realistic materials including
spin--orbit coupling, and the reciprocal current-induced phonon angular
momentum expected from Onsager reciprocity~\cite{lsv4-z14s} are left for
future work.

\section{Acknowledgments}
The authors thank H.-W. Lee, K.-J. Lee, D.~Go, J.~H. Han, H.-W. Ko, and S.~Han for helpful discussions.
This work was financially supported by the National Research Foundation of Korea (NRF) grant funded by the Korea government (MSIT) (RS-2024-00334933, RS-2024-00410027, RS-2026-25543309).


\section{Appendix A: Tight-binding model for numerical simulations}
\begin{figure}
	\centering
	\includegraphics[width=\linewidth]{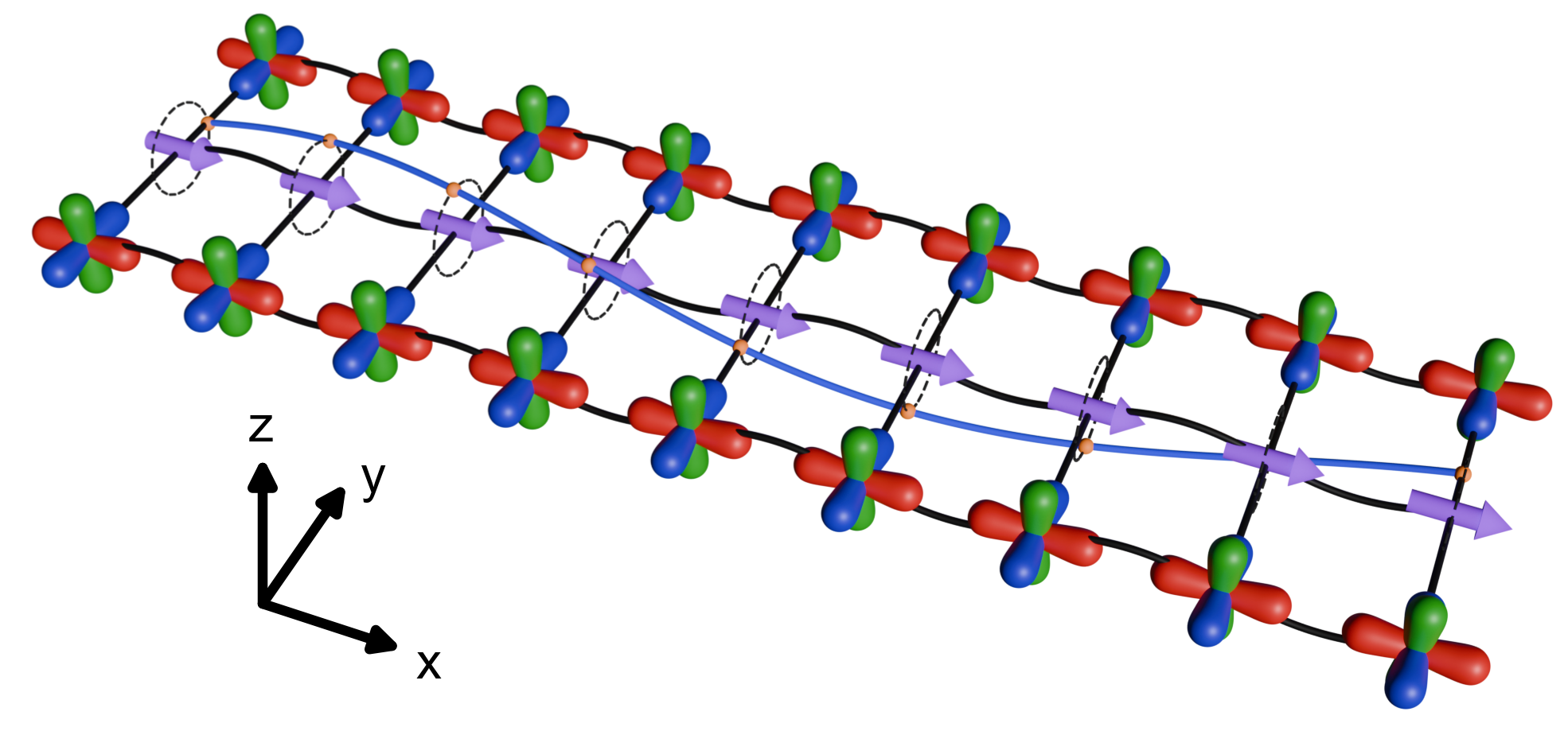}
	\caption{The $p$-orbital quasi-two-dimensional lattice model used in the
		simulations.}
	\label{fig:model}
\end{figure}

We use a quasi-two-dimensional tight-binding model with $p_x$, $p_y$, and
$p_z$ orbitals on a square lattice (Fig.~\ref{fig:model}). In the
Slater--Koster form, the hopping along the bond direction $\vhat\mu$ is
expressed through the projector $\vhat{\mu}\vhat{\mu}^T=I_3-L_\mu^2$, where $\vec{L}$ is the orbital angular momentum operator (divided by $\hbar$). This operator picks out the $p$-orbital lobe aligned with the bond (a $\sigma$ bond of amplitude $t_\sigma$) with the eigenvalue $1$, while the two perpendicular lobes form $\pi$ bonds of amplitude $t_\pi$, with the eigenvalue 0, and thus gives rise to the orbital
texture by the mechanism described in Refs.~\cite{goIntrinsicSpinOrbital2018,joGiganticIntrinsicOrbital2018,han2023}.
\begin{widetext}
	\begin{equation}
		H^{(0)}=\sum_{ij}\left\{c^\dagger_{i+1,j}
		\left[t_\pi+(t_\sigma-t_\pi)(I_3-L_x^2)\right]c_{ij}
		+c^\dagger_{i,j+1}\left[t_\pi+(t_\sigma-t_\pi)(I_3-L_y^2)\right]c_{ij}\right\}
		+\text{h.c.},
	\end{equation}
	A phonon with nonzero $\mathbf q$ changes the relative atomic positions
	$\Delta\mathbf d_{ij}$ between neighboring sites, and thereby the hopping
	integrals~\cite{mitraElectronphononInteractionModified1969,hamada2020PRR,hanOrbitalPumpingIncorporating2025a,yaoDynamicalOrbitalAngular2025a}. The transverse phonon correction responsible for orbital mixing is 
	\begin{equation}
		H^{(1a)}=-\frac{t_\sigma-t_\pi}{a}\sum_{ij}
		\left(c^\dagger_{i+1,j}\{L_x,\,\mathbf L\!\cdot\!
		\Delta\mathbf d^{x\perp}_{ij}\}c_{ij}
		+c^\dagger_{i,j+1}\{L_y,\,\mathbf L\!\cdot\!
		\Delta\mathbf d^{y\perp}_{ij}\}c_{ij}\right)+\text{h.c.},
	\end{equation}
\end{widetext}
with the displacement $\mathbf u_{ij}(t)=u_0[\cos(q_x x-\omega t)\vhat e_1
+\sin(q_x x-\omega t)\vhat e_2]$ and bond differences given by
$\Delta\mathbf d_{ij}^x=\mathbf u_{i+1,j}-\mathbf u_{ij}$ and $\Delta\mathbf d_{ij}^y=\mathbf u_{i,j+1}-\mathbf u_{ij}$, so that a uniform
displacement ($q=0$) cancels and produces no response. The superscript $\perp$ refers to the bond-perpendicular component. The anticommutator $\{L_\mu,\mathbf L\cdot\Delta\mathbf d^{\perp}\}$ shows that a transverse phonon tilts the hopping and mixes different orbital angular-momentum components, which is the microscopic origin of the OAM response. A longitudinal bond-stretch correction only modulates $(I_3-L_\mu^2)$ and produces no orbital mixing. The derivations of the transverse and longitudinal corrections, together with the numerical parameters used in the simulations, are provided in the Supplemental Material~\cite{supple}.

The Fermi level is set in a regime where $p_x,p_y$ dominate the Fermi-level sector and $p_z$ participates weakly. In this implementation the phonon wave vector is quantized by the finite lattice size, and the uniform-translation limit $q\to0$, where $\Delta\mathbf d_{ij}=0$, cannot be accessed directly. This is a limitation of the model, but it is consistent with the OAM response arising solely from an internal finite-$q$ deformation. Because the mechanism relies only on the projection structure $\vhat\mu\vhat\mu^T=I_3-L_\mu^2$, a $3\times3$ (three-orbital) structure, it applies to $p$ orbitals and similarly to $t_{2g}$ $d$ orbitals~\cite{chaudharyGiantEffectiveMagnetic2024}, with higher-multiplet cases following from a straightforward generalization~\cite{Han2022}.

In the present tight-binding mechanism, the phonon polarization determines which orbital-mixing vertices are activated through the combination of the bond direction and the transverse displacement. For $\vec{q}\parallel \vhat{x}$, a $y$-polarized displacement activates the $\{L_x,L_y\}$ vertex and directly mixes $p_x$ and $p_y$, whereas a $z$-polarized displacement activates $\{L_x,L_z\}$ and mixes $p_x$ and $p_z$. Thus, an $xy$-polarized circular phonon directly drives the $p_x$-$p_y$ channel, while a $yz$-polarized circular phonon coherently activates both $p_x$-$p_y$ and $p_x$-$p_z$ channels. Their combined dynamics can generate $p_y$-$p_z$ coherence and hence an $L_x$ response.

However, the AC orbital 
response need not coincide with the local-OAM channel: through the
velocity vertex and the interband orbital texture, the $yz$ phonon induces 
finite $L_y$ and $L_z$ AC components (see Fig.~\ref{fig:ACqw}). Thus, the circular displacement plane selects the directly activated orbital-mixing channels, while the observed AC/DC orbital components are set by the full response tensor.



\newpage


\begin{thebibliography}{99}
\bibitem{footnote1}
The transformation also leaves a second-order correction $\Delta H=-[(\vec u\cdot\nabla)\vec u]\cdot\nabla V$ to the crystal potential, which for a single-$q$ phonon averages to a uniform translation with no effect on bulk observables, as shown in Ref.~\cite{supple}.
\bibitem{footnote2}
Second-order contributions can also contain a $2\omega$ component, but we focus on the DC contribution because the $2\omega$ response is generally negligible compared with the leading AC $1\omega$ response.
\end{thebibliography}
\end{document}